\documentclass[a4paper]{jpconf}

\usepackage{glas}

\usepackage{graphicx}
\usepackage{amssymb}
\usepackage{epstopdf}
\begin{document}

\begin{titlepage}{GLAS-PPE/2010-??}{14$^{\underline{\rm{th}}}$ February 2011}

\title{Establishing Applicability of SSDs to LHC Tier-2 Hardware Configuration}

\author{Samuel C Skipsey$^1$, Wahid Bhimji$^2$, Mike Kenyon$^{3}$\\
\\
$^1$ School of Physics and Astronomy, University of Glasgow, Glasgow G12 8QQ, United Kingdom\\
$^2$ University of Edinburgh, School of Physics \& Astronomy, Edinburgh EH9 3JZ, United Kingdom\\
$^3$ IT Department, CERN, Route de Meyrin 1121, Geneva 23, Switzerland}

%

\begin{abstract}
Solid State Disk technologies are increasingly replacing high-speed hard disks as the storage technology in high-random-I/O environments. There are several potentially I/O bound services within the typical LHC Tier-2 - in the back-end, with the trend towards many-core architectures continuing, worker nodes running many single-threaded jobs and storage nodes delivering many simultaneous files can both exhibit I/O limited efficiency. We estimate the effectiveness of affordable SSDs in the context of worker nodes, on a large Tier-2 production setup using both low level tools and real LHC I/O intensive data analysis jobs comparing and contrasting with high performance spinning disk based solutions. We consider the applicability of each solution in the context of its price/performance metrics, with an eye on the pragmatic issues facing Tier-2 provision and upgrades

\vspace{0.5cm}
\begin{center}
{\em International Conference on Computing in High Energy and Nuclear Physics (CHEP) 2010}\\
{\em Taipei, Taiwan}
\end{center}
\end{abstract}

\newpage
\end{titlepage}

\section{Introduction}
Trends in CPU manufacture over the last several years suggest that the multi-core architecture of modern chips is only going to increase over the short-term future \cite{manycore}. Current Particle physics codes are effectively single-threaded, and sites within WLCG\footnote{Worldwide LHC\footnotemark{} Compute Grid}\footnotetext{Large Hadron Collider} are currently recommended to provide one job slot in their batch system per core (or even, for hyperthreaded processors, one per virtual core\cite{andypaper}). This results in a large number of processes (on the order of 8 to 12, for commonly used quad,hex-core, dual-processor configurations\footnote{Even at the present day, this number could be as high as 48, for a duodec-core, dual-processor, or hex-core, quad-processor configuration, both of which are available.}), each attempting to perform I/O operations on the system storage device.
For the workloads involved in monte-carlo production, this additional load is almost insignificant to job efficiency, as such jobs are primarily CPU-bound. For analysis workloads, however, I/O is the limiting factor\cite{analysisSTEP}, and thus the I/O contention created by concurrent analysis jobs on a single worker node is of significance to job efficiency (and reliability).
Much of this I/O load is caused by seeks within the ROOT\cite{rootpaper} file structures used by particle physics to store event data \cite{wahidpaper}.

SSDs\footnote{Solid State Disks} are an emerging technology which seeks to compete with conventional ``spinning-disk'' magnetic hard-disk technologies (``HDDs'') across almost all applications. SSDs are usually based upon the ubiquitous Flash memory technology. There are two types of Flash memory, with separate applications based on their cost/capacity ratio and performance. SLC Flash is more expensive per GB, because it only stores a single value per cell, but is more performant for the same reason. MLC Flash stores multiple values per cell, allowing it a higher storage density, and lower cost/capacity ratio. As a result, MLC Flash is more often found in commodity SSD devices aimed at HDD replacement in desktop computer and laptops, and SLC Flash in high-performance devices aimed at enterprise applications (like metadata storage, or database acceleration).

As Flash storage is bus-addressable, there is little additional seek-time incurred in accessing data out-of-order from the storage array (no physical parts need move, and the only overhead is in skipping the contents of the onboard DRAM readahead buffer), especially compared to HDDs, which must wait for the physical media to rotate under the read-head to reach any out-of-order data\footnote{Native Command Queuing allows an HDD controller to buffer requests for out-of-order data for a certain period, in order that it might reorder the requests into a more optimal sequence on the physical disk medium. This can reduce the effective latency for very `seeky' loads, but it is necessarily limited in the extent to which it can be applied.} The potential application of SSDs to the analysis I/O issue is clear: if the problem for multiple jobs is seeks, why not use a technology where seeks are fast?

\subsection{ROOT file optimisation}
Another approach to addressing the I/O load issues is to attempt to improve the quality of the ROOT files themselves. If accesses within a ROOT file can be made more in-order (more ``linear''), then each process' I/O looks more like a streaming process, with the only seeks being when switching context between the I/O requests of different processes. In fact, all of the WLCG experiments have embarked on improvements to their file formats in this manner\cite{atlasreorder,cmsreorder}. We will consider the effect on SSD applications later in this paper, but the actual changes involved are outside of our scope.
 
\section{Context and methodology}
We evaluated the performance improvements in standard worker nodes in the University of Glasgow  compute cluster allocated for GridPP for various potential replacements for the conventional HDD.

Two types of SSD, both of them using MLC Flash, were considered. No SLC Flash based SSD was evaluated for this application as the prices for suitable devices were outside the acceptable range for a Tier-2 site's procurement funding (and therefore, pragmatically, there is no point in testing them). A previous piece of research\cite{SSDref} has already been performed by others, evaluating the performance of SLC Flash modules for similar analyses. They show the expected significant performance increases over HDDs, although the authors note that the price and capacity of such units render them impractical.

Of the two SSDs used, the Kingston SSD Live! V-series 128GB is a low-end consumer-grade solution intended for laptop use. We use this as an indicator of the bottom of the budget performance possible from an SSD.

The second SSD type used was the Intel SSD X-25M G2 160GB. Intel SSDs are generally regarded\cite{IntelSSDreview} as solid performers in both the consumer and enterprise SSD environments, and the G2 (second generation) models in particular are regarded as having very good price/performance for their class. These SSDs are intended to represent the mid-high end of the consumer SSD market, and also the high end of the affordable price range for a Tier-2 site.

\subsection{blktrace and I/O analysis}
The low-level I/O capture tool, blktrace\cite{blktrace}, was used to capture live I/O patterns on each of the test configurations for later analysis. Blktrace makes use of the kernel debug filesystem to extract data about I/O activity at the kernel level in a straightforward way. As such, it is only usable on distributions supporting kernel releases post-2.6.10,  a set which excludes RHEL4 (and derivatives, like Scientific Linux 4), but includes RHEL5 (and thus SL5). As part of the intent of the work was to examine the behaviour of `real systems', no attempt was made to backport more recent kernels to SL4 systems. As a result, all of our monitoring is on then-current SL5 systems running the then-current releases of the gLite middleware.

Blktrace output is parsed with the open-source Seekwatcher tool, which produces graphical representations of the I/O pattern, and the seek rates over the sampled period.

\subsection{HammerCloud framework}
The HammerCloud\cite{analysisSTEP} automated job submission and testing framework was used to manage job load and type for all tests performed for this paper.

In particular, the efficiencies for the primary results presented are derived from HammerCloud test runs 1332, 1334 and 1348. 

\subsection{Methodology}
Jobs were constrained to run only on a subset of the batch system, which was reserved for the user accounts that HammerCloud test jobs are mapped to. As a result, these nodes only executed jobs provided by the test infrastructure. The results marked ``HDD'' are for a conventional worker node configuration, with a single 7200RPM hard disk partitioned for system and tmp filesystems. Our configuration creates scratch space for incoming jobs in the tmp filesystem, so all significant I/O due to user workloads is located within this filesystem.
For the SSD rows, the tmp filesystem, only, was mounted on the relevant SSD. Monitoring data for blktrace was written to the system disk (the conventional hard disk), to avoid interfering with the I/O for the jobs.
For the RAID rows, the nodes were prepared logically as for the ``HDD'' configuration, with one storage volume partitioned into system and tmp filesystems. However, the storage volume was either a RAID 1 or RAID 0 array created in software from two hard disks identical to the single disk found in the conventional configuration.

\section{Results}

Table \ref{primaryresults} summarises the result of the performance tests for HammerCloud tests using the FileStager data access method, and ATLAS reordered AODs. 
\begin{table}
\caption{\label{primaryresults}Summary of job efficiencies from HammerCloud FileStager test loading with various underlying storage.}
\begin{center}
\begin{tabular}{llll}
\br
Cores:Jobs&Storage Type&Mean Job Efficiency&Mean Throughput\\
\mr
\multicolumn{4}{c}{Conventional 8-core Node}\\
\mr
8:8&Kingston Value SSD&60\%&4.5\\
8:8&HDD&75\%&5.5\\
8:8&Intel X-25 SSD&80\%&6\\
8:8&2$\times$HDD RAID 1&83\%&6.6\\
8:8&2$\times$HDD RAID 0&90\%&7\\
\mr
\multicolumn{4}{c}{Magny-Cours 24-core Node}\\
\mr
24:24&Intel X-25 SSD&50\%&12\\
24:24&2$\times$HDD RAID 0&86\%&21\\
\mr
\multicolumn{4}{c}{Single Job Efficiency (Measured)}\\
\mr
8:1&HDD&90\%&0.9\\
\br
\end{tabular}
\end{center}
\end{table}

\begin{figure}[h]
\begin{minipage}{20pc}
\includegraphics[width=20pc]{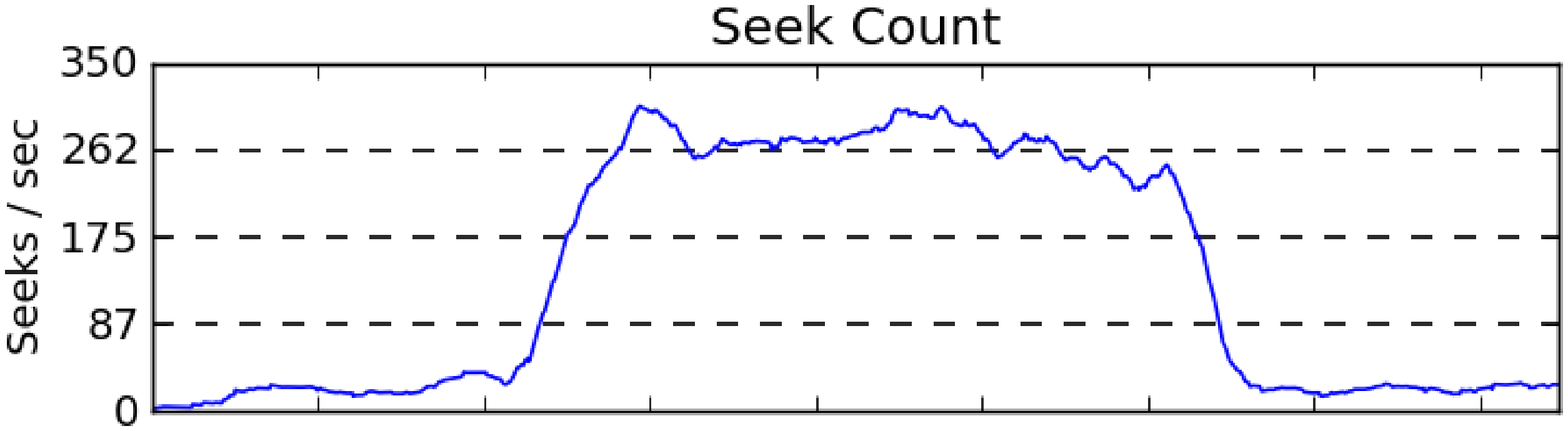}
a) unordered AODs
\end{minipage}
\begin{minipage}{20pc}
\includegraphics[width=20pc]{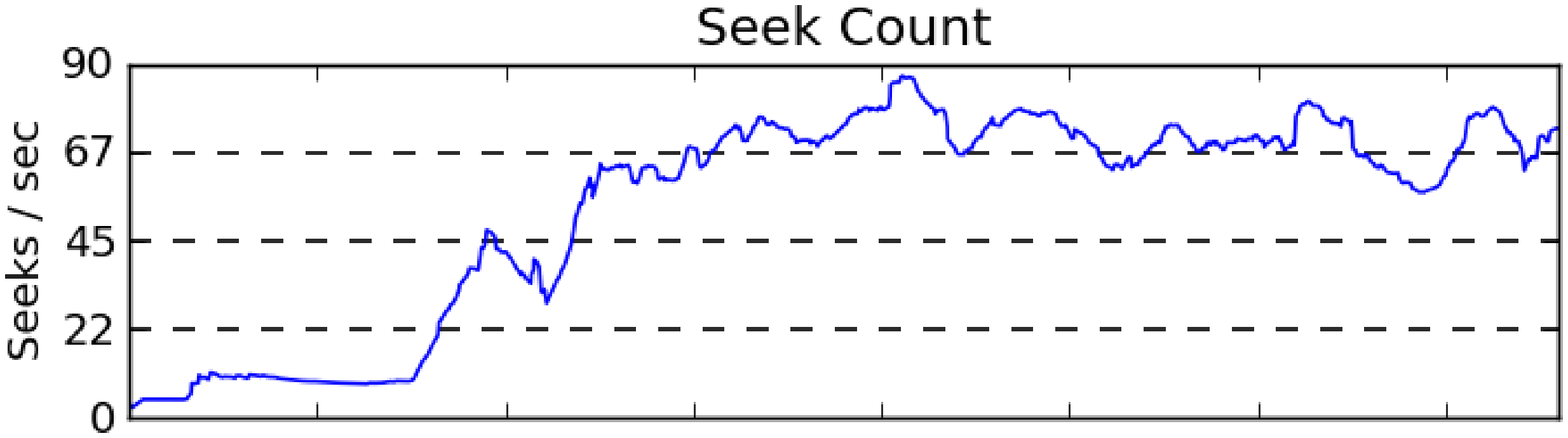}
b) reordered AODs
\end{minipage}
\caption{\label{unorderedvsordered}Blktrace-gathered I/O record for multi-job access to `unordered' and `reordered' AODs, generated using the Seekwatcher tool.}
\end{figure}

The difference between reordered and unordered AOD file access on these results can be seen from Figure \ref{unorderedvsordered}. As can be seen, the improved ordering process on the serialised trees in the `reordered' AODs significantly reduces the peak seek rate when accessing the files on an SSD. Preliminary results against unordered AODs on SSDs showed SSD performance in a better light than with the reordered AODs now in use; however, we do not include these results as they are irrelevant to the current state of play.

Another possibility for removing local I/O contention is to use remote file I/O interfaces to access files directly from the site local storage element, rather than staging them to the worker node storage. In ATLAS use, this is implemented by the DQ2\_LOCAL data access method. Table \ref{rfioresults} shows efficiencies for the same configuration as table \ref{primaryresults}, but using the DQ2\_LOCAL method. As can be seen, none of the efficiencies reach the 90\% efficiency seen for single jobs (or 8 jobs on a 2$\times$HDD RAID 0 configuration). However, for the specific instance of the Kingston Value SSD, performance is measurably improved by removing the filestaging process from the workload. In all other cases, performance is lower than for the equivalent case with locally staged files.
For the Magny-Cours node, there is one caveat for these results; a node with this degree of job parallelism would almost certainly be provided with a 10GigE, or equivalent, interface, rather than the 1GigE interface used for this test. Thus, the actual efficiencies for a ``real'' high-parallelism node with remote I/O will have these results as a lower-bound.

\begin{table}
\caption{\label{rfioresults}Summary of job efficiencies from HammerCloud DQ2\_LOCAL test loading with various underlying storage.}
\begin{center}
\begin{tabular}{llll}
\br
Cores:Jobs&Storage Type&Mean Job Efficiency&Mean Throughput\\
\mr
\multicolumn{4}{c}{Conventional 8-core Node}\\
\mr
8:8&Kingston Value SSD&67\%&5.35\\
8:8&Intel X-25 SSD&73\%&5.86\\
8:8&2$\times$HDD RAID 1&73\%&5.88\\
8:8&HDD&78\%&6.25\\
\mr
\multicolumn{4}{c}{Magny-Cours 24-core Node}\\
\mr
24:24&2$\times$HDD RAID 0&73\%&17.4\\
24:24&3$\times$HDD RAID 0&76\%&18.2\\
\br
\end{tabular}
\end{center}
\end{table}

After the conclusion of the first phase of testing, we concluded that the advantage of RAID-0 disk configurations was proved enough to invest in retrofitting the cluster with such a configuration. For long-term testing reasons, we retained the Intel X-25 SSDs in those nodes fitted with them, as the performance deficit was considered significant only when the node was fully loaded with analysis work (and no cores performing other workloads). It should be noted that, during the same time period, the ATLAS file-caching utility, pCache\cite{pCache}, was also installed on the cluster, which would have reduced the number of writes needed to the filesystems to some degree\cite{gridpppCache}. Table \ref{longtermtest} shows the mean efficiencies for the RAID 0 and SSD nodes when full of real ATLAS analysis work. As can be seen, the `real world' values are within a few percent of the simulated values provided by the HammerCloud test. These data points were taken for the first week in September 2010, filtering out periods when the nodes sampled were not fully occupied.

\begin{table}
\caption{\label{longtermtest}Summary of job efficiencies for real ATLAS analysis workloads (taken from periods when the nodes were uniformly filled) on 8-core worker nodes.}
\begin{center}
\begin{tabular}{ll}
\br
Storage Type&Mean Job Efficiency\\
\mr
Intel X-25 SSD&78\%\\
2$\times$HDD RAID 0&88\%\\
\br
\end{tabular}
\end{center}
\end{table}

It may be useful to put these results in context, by casting them in terms of the price/performance ratio for two metrics of performance for storage - the throughput achieved in the total worker-node system (from Table \ref{primaryresults}), and the capacity of the storage volume (which is particularly important as the amount of data processed by WLCG VO analysis is increasing over time). Table \ref{priceperform} shows these figures for the three test cases we examined. As can be seen, the price/performance ratios for these `real world' metrics are much worse for the SSDs than they are for the simple RAID0 array, by around a factor of 3 to 10. Of course, as a fraction of the cost of a complete worker node, these costs are relatively smaller (perhaps 10\% of the total system cost), but the ratios of the metrics remain the same.

\begin{table}
\caption{\label{priceperform}Price/performance metrics for the SSDs used, compared to a conventional harddisk}
\begin{center}
\begin{tabular}{llll}
\br
Storage Type&Unit Price(GBP)&Price/Throughput(8core node,GBP)&Price/GB(GBP)\\
\mr
Kingston SSDnow&155&34.40&1.21\\
Intel X-25 SSD&245&40.80&1.53\\
2xHDD(500GB)&80&11.40&0.16\\
\br
\end{tabular}
\end{center}
\end{table}


\section{Discussion}

Although the popular view of SSDs is that of an expensive, but always more performant, replacement for the hard disk drive, our results show that the facts do not always bear this out. 
In particular, the performance of affordable SSDs is dramatically weaker whenever widespread writes as well as reads are required, especially when the number of seeks required is relatively small. It is notable, however, that increasing the number of jobs acting on a node (and thus the number of seeks switching between I/O threads) does not increase the relative performance of the SSD in our testing.

Regardless of the fact that such SSDs are pragmatically out-performed by RAID0 arrays for the workloads for which Tier-2 sites are concerned, it is further important to consider their performance with respect to the price spent on them. The significant costs of an SSD of a reasonable capacity compared to commodity 7200 RPM disks worsen their price/performance metrics significantly in all cases; even against a lone 7200RPM disk, their metrics are poor.

\section{Conclusions and Further Work}
In conclusion, we find that for the average Tier-2 site, where price must also be a factor in purchasing decisions, SSDs are not a viable technology for worker node internal storage. 
It is possible that more advanced approaches to worker node storage use (provisioning of a cache for file transfers, from which symlinks could be made for instances of repeatedly requested files on the same node) would provide a more natural niche for SSD use, by increasing the ratio of reads to writes on the storage. Unfortunately, the most mature such technology in use by a VO, pCache, is incapable of using separate filesystems for cache and job scratch space, as it uses hardlinks as an efficient means of reference counting for cache management. Additionally, this space is perhaps more naturally occupied by the more performant distributed or network filesystems - Lustre, GPFS, pNFS, and the like, as this allows multiple workers to access the same cache, increasing the chance of hits on any particular request. The use of SSDs as metadata storage targets for Lustre is already known to be useful, but it is outside of the scope of this investigation.

\ack

The authors would like to thank Dan van der Ster (and Johannes Elmsheuser) for providing the HammerCloud framework necessary for this research, and allowing the large number of rapid tests which we submitted with it.

\end{document}